\documentclass{aastex63}

\graphicspath{{./}{figures/}}

\begin{document}

\title{Nanoflare Theory Revisited}

\correspondingauthor{Amir Jafari}
\email{elenceq@jhu.edu}

\author{Amir Jafari}
\email{elenceq@jhu.edu}
\affiliation{Department of Physics \& Astronomy, Johns Hopkins University, Baltimore, USA}

\author{Ethan T. Vishniac}
\email{evishni1@jhu.edu}
\affiliation{Department of Physics \& Astronomy, Johns Hopkins University, Baltimore, USA}

\author{Siyao Xu}
\email{sxu93@wisc.edu}
\affiliation{University of Wisconsin, Madison, USA}








\begin{abstract}
Local magnetic reversals are an inseparable part of magnetohydrodynamic (MHD) turbulence whose collective outcome on an arbitrary scale in the inertial range may lead to a global stochastic reconnection event with a rate independent of small scale physics. In this paper, we show that this picture is intimately related to the nanoflare theory proposed a long time ago to explain the solar coronal heating. First, we argue that due to stochastic flux freezing, a generalized version of flux freezing in turbulence, the magnetic field follows the turbulent flow in a statistical sense. Bending and stretching an initially smooth field, therefore, the turbulence generally increases the magnetic spatial complexity---a measure of the geometric complexity of the field recently formulated in terms of renormalized field at different scales. Strong magnetic shears associated with such a highly tangled field can trigger local reversals and field annihilations that convert magnetic energy into kinetic and thermal energy respectively. The former maintains the turbulence, which incidentally continues to entangle the field completing the cycle, while the latter enhances the heat generation in the dissipative range. We support this theoretical picture invoking recent analytical and numerical studies which suggest a correlation between magnetic complexity and magnetic energy dissipation. The amplification of multiple local, in-phase reversals by super-linear Richardson diffusion may initiate a global reconnection at larger scales, however, even in the absence of such a global stochastic reconnection, the small scale reversals will continue to interact with the turbulence. We employ conventional scaling laws of MHD turbulence to illustrate that these local events are indeed efficient in both enhancing the turbulence and generating heat. Finally, using an MHD numerical simulation, we show that the time evolution of the magnetic complexity is statistically correlated with the kinetic energy injection rate and/or magnetic-to-thermal energy conversion rate.

\end{abstract}



\section{Introduction}

Over half a century ago, \cite{Grotrian1939} and \cite{Edlen1942} pointed out that the unexpected emission lines detected in the spectrum of the solar corona indicated a very high ionization, which would require extremely high temperatures of order $10^6$ K. One implication is that the solar corona is much hotter than the lower layers, which are much closer to the sun. This is puzzling, of course, because one expects the temperature to drop off with distance from the solar surface. Several mechanisms have been proposed to resolve this theoretical difficulty, however, it is still the subject of ongoing research. Interestingly, though, almost all of these models rely, in one way or another, on a common phenomenon---magnetic fields.

Alfv\'en wave dissipation and magnetohydrodynamic (MHD) shocks \citep{Moriyasuetal.2004, McIntoshetal.2011}, magnetic reconnection \citep{Roaldetal.2000, Liuetal.2002, Aulanieretal.2007, Hoodetal.2009} and MHD turbulence \citep{Cranmeretal.2007, Rappazzoetal.2007} are among the proposed mechanisms to solve the coronal heating problem as it is called now. Each model is often backed up by few numerical simulations, as is usual nowadays, yet there is no consensus on which processes play a more fundamental role. A typical model identifies a source of energy, e.g., magnetic energy or energy carried by Alfv\'en waves, which is finally converted into thermal energy. The second component is a mechanism to do the energy conversion, e.g., MHD turbulence damping Alfv\'en waves which generates heat. It is highly plausible, on the other hand, to think that several mechanisms work together to give rise to such a bizarre situation. In any case, one should keep in mind that real astrophysical systems, including the solar corona, are much more complicated than what a simple theoretical picture may present based on few physical mechanisms. In any case, magnetic fields seem to play a very important role in theories that attempt to explain the coronal heating problem
\citep{Golub1980, HeyPri1984}.

As one of the mechanisms possibly responsible for, or else partly contributing to, the solar coronal heating phenomenon, magnetic reconnection has been frequently invoked through both analytical and numerical studies.
Recent advances in both observations, with high spatial and temporal resolution, and numerical simulations in studying the solar coronal heating problem are summarized by \cite{Moor2015}. See also reviews by e.g., 
\citealt{Priest, Low2003}
on the role of magnetic reconnection in the coronal heating and solar coronal phenomena
and reviews by e.g., 
\cite{Ulms1991, Spicer1991, Somov1994, Parnell2012}
for different coronal heating mechanisms. The fragmented and turbulent nature of magnetic reconnection has been confirmed by observations of a super-hot current sheet during SOL2017-09-10T X8.2-class solar flare 
\citep{Cheng2018}. \cite{Long2015}
quantified the contribution of magnetic reconnection to the coronal heating for one special case of an active region (AR 11112) by measuring the rate of magnetic reconnection and the rate of energy dissipation in the solar corona. Extrapolating the result to other regions, they concluded that magnetic reconnection can in fact account for the measured temperatures. By comparing with extreme-ultraviolet observations, \cite{Yang2018}
argued that the impulsive reconnection is responsible for the active region coronal heating. In terms of numerical simulations, \cite{Kanella2017} identified individual heating events in 3D MHD simulations of the solar corona and the corresponding released energy rate and volume ranges. Their results suggest the stochastic nature of magnetic reconnection in releasing a random fraction of the energy stored in the magnetic fields as heat. 
The kinetic particle-in-cell (PIC) simulations performed by \cite{Shay2018}, to study the heating effects of magnetic reconnection, showed that the statistics of the turbulent reconnection is important for determining the ion and electron heating. In a recent review, 
\cite{Vlah2019} provided evidence from numerical simulations showing that the turbulent reconnection with spontaneous formation of current sheets in the solar corona drives both coronal heating and particle acceleration.

Magnetic reconnection generates fast, explosive motions in magnetized fluids thereby enhancing diffusion at large scales. It may also start a turbulent cascade in an initially quiet medium or else it may help the present turbulence by injecting kinetic energy at large scales, typically of order the scale of the reconnection zone
\citep{Kowal2017}. In the stochastic model of reconnection \citep{LazarianandVishniac1999}, energy is basically injected on a range of scales, which provides a more efficient way to enhance turbulence as we will show in the present paper. On the other hand, it is well known that resistivity is not enough to generate appreciable heat in typical astrophysical systems. True, the Sweet-Parker model \citep{Parker1957, Sweet1958} predicts much faster magnetic energy conversion rate than the magnetic dissipation, however, it is still very slow \citep{Yamadaetal.2010, JV2018Review}. Although a small, but finite, resistivity is required for stochastic reconnection to start and proceed, but neither the reconnection rate nor its underlying mechanism does depend on resistivity. The major role, instead, is played by the turbulence. In typical reconnection models, a global reversal converts  magnetic energy into kinetic energy and pumps it into the medium at large scales. This will generally enhance diffusion at large scales, not necessarily generating a fully developed turbulence. In addition, such large scale motions are generated only in occasional global reconnections. In contrast, local stochastic reconnections over a continuum of scales can generate small scale motions efficiently enhancing the local turbulent cascade even in the absence of a global field reversal. This constitutes a more effective way of generating, enhancing and maintaining turbulence. As discussed before, the topological dissipation of stochastic magnetic fields has been identified as an alternative mechanism of coronal heating
\citep{Parker1972, Levine1974}. The dissipation of magnetic energy via magnetic reconnection occurs at many small-scale tangential discontinuities (current sheets), which are caused by the photospheric footpoint motions 
\citep{Parker1987}. These ubiquitous impulsive heating events are referred to as nanoflares. \cite{Parker1988} also suggested that the initially slow reconnection can be enhanced by hydromagnetic and plasma turbulence and thus has a later explosive reconnection phase. 
The nanoflare model for coronal heating has been further investigated both analytically and numerically by different authors, see e.g., \cite{Carg2004, Rapp2008, Parn2010, Bown2013, Jess2019}.

In the present paper, using both analytical and numerical considerations, we argue that local reversals in MHD turbulence can be studied in a statistical framework which unifies reconnection and other magnetic phenomena, in particular, the magnetic heating process invoked in the nanoflare theory. We illustrate that stochastic reconnection, as the global outcome of many simultaneous local reversals, is more efficient than conventional models in enhancing turbulence and heat generation. Finally, we show that even in the absence of a global reconnection at larger scales of order the system's scale, the local reversals at comparatively smaller scales in the inertial range can maintain and enhance the turbulence and also the process of heat generation at much smaller scales in the dissipative range. The theoretical picture which we invoke to advance our arguments can be briefed as follows: Turbulence stretches and bends the magnetic field, which follows the turbulent flow in a statistical sense due to stochastic flux freezing \citep{Eyink2011}, producing magnetic gradients at random regions in the turbulence inertial range. The resultant magnetic shears can give rise to local, small scale magnetic reconnection events whose collective outcomes may lead to a global reconnection event \citep{LazarianandVishniac1999, Review2020}. As the magnetic field gets stretched, bent and tangled by the turbulence, its spatial complexity increases in a geometric sense \citep{JV2019}. The field resists tangling because of the magnetic tension forces, which at some point make the field slip through the fluid to relax \citep{JV2019, Jafari2020}. The relaxing field may in turn accelerate particles \citep{DeG05, Kow12, Khia2015, Bere2016,Lu2020}
producing jets of fluid. Hence, during reconnection, the spatial complexity of the velocity field increases while that of magnetic field decreases after reaching a local maximum. Previous work has in fact quantified the level of spatial complexity associated with  magnetic and velocity fields \citep{JV2019, JVV2019, Jafari2020} with the implication that these local reconnection events are ubiquitous in MHD turbulence. This picture in fact reformulates the theory of coronal heating by local nanoflares, proposed a long time ago by \cite{Parker1972} to explain the solar coronal heating, in a statistical picture connecting it to the stochastic model of magnetic reconnection \citep{LazarianandVishniac1999}.

As for the detailed plan of this paper, we start off by revisiting dissipative anomalies and stochastic reconnection in \S\ref{sturbulence} to illustrate how magnetic flux freezing fails in turbulence, which is intimately related to stochastic reconnection. These considerations are already well-established and our emphasis here is due to the fact that they play a major role in the development of the main ideas of this paper. For a detailed review of stochastic reconnection and stochastic flux freezing, see e.g., \cite{JV2018Review, Review2019, Review2020}. In \S\ref{SHeating}, which presents the main results of this paper, we use the notion of vector field complexity to argue that local reversals involved in stochastic reconnection are efficient in both maintaining the local turbulent cascade and also heating the fluid. To support these statistical arguments, we also use simple scaling laws in MHD turbulence, which are model-independent in general although we use the Goldreich-Sridhar model \citep{GoldreichandSridhar1995, GoldreichandSridhar1997} to illustrate the main points. Furthermore, we also test our main results using an incompressible, homogeneous MHD turbulence numerical simulation. Finally, in \S\ref{Discussion}, we summarize and discuss our results.

\section{MHD Turbulence}\label{sturbulence}
In this section, we present a brief review of the tools required to study stochastic heating in turbulent fluids, including dissipative anomalies in incompressible fluids \citep{Eyink2018, JV2019}, which make the turbulent velocity field H{\"o}lder singular (see below), and the failure of flux freezing in MHD turbulence \citep{Eyink2011, Eyinketal2013}. We also briefly revisit stochastic magnetic reconnection \citep{LazarianandVishniac1999, JV2018Review, Review2019, Review2020}.

\subsection{Dissipative Anomalies}\label{sturbulence1}

In a magnetized fluid with a large characteristic length scale $L$, or a large characteristic velocity $U$, or a tiny viscosity $\nu$, the Reynolds number $Re=LU/\nu$ can be very large. If the magnetic diffusivity $\eta$ is of the same order as the viscosity $\nu$, implying a magnetic Prandtl number of order unity $Pr_m=\nu/\eta\sim 1$, the magnetic Reynolds number $Re_m=LU/\eta$ will be large too. In order to see what a large kinetic Reynolds number means, we can re-write the momentum (Navier-Stokes) equation, in the common notation, using the parameters

$${\overline {\bf x}}={\bf x}/L,\;\;\;{\overline t}=t/(L/U),\;\;\;{\overline{\bf u}}={\bf u}/U,\;\;\;{\overline p}=p/U^2,$$

in a dimensionless form as 

$${\partial \overline {\bf u}\over \partial \overline t}+\overline{\bf u}.\overline\nabla\overline{\bf u}=-\overline\nabla\overline p+{1\over Re}\overline\nabla^2\overline{\bf u}.$$

Let us assume an incompressible fluid; $\nabla.{\bf u}=0$. Apparently, as $Re=LU/\nu$ increases, by either increasing the system's characteristic size or velocity or decreasing the viscosity, the last term in the momentum equation tends to vanish. This might for example justify ignoring a small viscosity altogether in some cases, but not always. As the Reynolds number increases, i.e., $Re\rightarrow \infty$, the flow becomes unstable: like a pen balanced on its tip, any small perturbation would lead to turbulence \citep{JV2019, Jafari2020}. This is why the initially slow and laminar flow coming out of a faucet would become turbulent at some point if we keep increasing the flow velocity $U$ (i.e., increasing $Re$). Indeed, large Reynolds numbers, frequently encountered in astrophysical fluids, are typically associated with turbulence. On the other hand, numerous numerical simulations and experiments have shown \citep{Sreenivasan1984, Sreenivasan1998, EyinkS2006, Eyink2018} that in turbulence the kinetic energy dissipation rate $\epsilon_k(t)=\nu|\nabla {\bf u}|^2$ does not approach zero when viscosity tends to vanish, rather it approaches a non-zero constant $\lim_{\nu\rightarrow 0}\nu|\nabla {\bf u}|^2\rightarrow \epsilon_k^*>0$---the phenomenon of dissipation anomaly. Thus the velocity gradients should diverge in the limit of vanishing viscosity, $\lim_{\nu\rightarrow 0} |\nabla\bf u|\rightarrow \infty$, to keep $\nu|\nabla {\bf u}|^2$ constant. With diverging and ill-defined velocity gradients, hydrodynamics equations will consequently become ill-defined in ideal turbulence; for more details see e.g., \cite{Eyink2018, JV2019}. Incidentally, in passing, we should note that the limit $\nu \rightarrow 0$ (or equivalently $Re\rightarrow \infty$) is just the mathematical translation of the physical statement that one can take an arbitrarily small viscosity (or an arbitrarily large $Re$): viscosity is not required, or assumed, to vanish---viscosity never vanishes but it can be taken as small as one wishes.

Similar to the momentum equation, the induction equation governing the evolution of magnetic field $\bf B$, with a characteristic strength $\cal B$, can be written in a dimensionless form as follows:

$${\partial \overline{\bf B}\over \partial \overline t}={1\over Re_m}{\overline \nabla}^2 \overline{\bf B}+{\overline\nabla}\times({\overline{\bf u}}\times\overline{\bf B}),$$

where $\overline{\bf B}={\bf B}/{\cal B}$. In turbulence, magnetic dissipation rate $\epsilon_m(t)=\eta|\nabla{\bf B}|^2$ does not approach zero as the diffusivity tends to vanish, i.e., $\lim_{\eta\rightarrow 0}\eta|\nabla{\bf B}|^2\nrightarrow 0$ (magnetic dissipation anomaly). Magnetic field gradients diverge, $|\nabla{\bf B}|\rightarrow \infty$, and MHD equations become ill-defined as a result; see e.g., \cite{Eyinketal2013, JV2019}. It is physically naive and mathematically incorrect, therefore, to ignore viscosity altogether and use ideal fluid equations in real fluids, unless we apply careful measures to keep the Reynolds number small to avoid the development of turbulence. Likewise, a vanishingly small magnetic diffusivity cannot justify ignoring the diffusivity altogether. If turbulence is developed, magnetic and velocity gradients will typically become ill-defined or singular\footnote{Mathematically, this means that these vector fields become H{\"o}lder singular instead of being Lipschitz continuous. For a Lipschitz function $f(x)$, the slope (derivative) at any point of the domain has an upper bound, i.e., there is a positive constant $f_L$ such that $|f(x_2)-f(x_1)|\leq f_L |x_2-x_1|^h$ with $h=1$. For H{\"o}lder functions $0<h<1$, which means that the slop can increase indefinitely. Generalization to vector fields is straightforward: the field ${\bf B(x)}$ satisfying $|| {\bf B(x)}-{\bf B(y)}  ||\leq B_0 |{\bf x-y}|^h$, with $B_0>0$, is Lipschitz continuous if $h=1$, and H{\"o}lder singular if $0<h<1$. In the latter case, $\nabla\bf B$ will in general become ill-defined.}, i.e., the field gradients will diverge. If we insist to use MHD equations, which we do, we would have to remove these singularities first. One way to do so is to smooth the fields or, in other words, to use the average velocity field ${\bf u}_l({\bf x}, t)$ or magnetic field ${\bf B}_l({\bf x}, t)$ in a parcel of fluid of length scale $l$ located at the spacetime point $({\bf x}, t)$ instead of using the bare, mathematical fields ${\bf u}({\bf x}, t)$ and ${\bf B}({\bf x}, t)$. This simple coarse-graining methodology, to be revisited in \S\ref{SHeating}, can be applied to any scalar or vector field in MHD turbulence \citep{JV2019}. 

\subsection{Failure of Flux Freezing}

One important implication of the above considerations, as far as the reconnection of turbulent magnetic fields is concerned, is the breakdown of the standard Alfv\'en flux-freezing law \citep{Alfven1942} in turbulent systems. If the flow remains laminar but the diffusivity $\eta$ is very small, under certain conditions, the diffusive term $\eta \nabla^2 {\bf B}$ may be ignored in the bare (i.e., not coarse-grained) induction equation, $D_t{\bf B}={\bf B.\nabla u-B\nabla. u}+\eta \nabla^2 {\bf B}$ with Lagrangian derivative $D_t\equiv (\partial_t+{\bf u.\nabla}$). Thus, using the continuity equation $D_t\rho+\rho\nabla.{\bf u}=0$, one finds $D_t\Big({ {\bf B}/ \rho}\Big)=\Big( {{\bf B}/ \rho}\Big).\nabla{\bf u}$, which means that the magnetic field is frozen into the fluid, i.e., the integral curves of ${\bf B}/\rho$ are advected with the fluid and the field follows particle trajectories. However, at least in most astrophysical systems, a vanishingly small diffusivity (i.e., a large $Re_m$) will typically be accompanied with a small viscosity\footnote{For instance, in highly ionized accretion disks, in which the magneto-rotational instability (MRI) is thought to be active, $Pr_m$ is usually assumed to be of order unity \citep{JV2018} while it is much smaller in planetary and stellar interiors. In any case, at least in astrophysics, huge kinetic Reynolds numbers are typically accompanied with huge magnetic Reynolds numbers.}, which translates into large kinetic Reynolds numbers, i.e., turbulence. Hence, the induction equation used in the above derivation of ideal flux freezing will not remain well-defined because of the blow-up of velocity and magnetic gradients. All other derivations of the Alfv\'en flux freezing law, in a similar way, assume that the induction equation (and/or other MHD equations) are well-defined thus neither of such derivations guarantees the validity of flux freezing in turbulence. Indeed the Alfv\'en flux freezing theorem fails in turbulence. Particle trajectories are random in turbulent flows, therefore, the magnetic field which tends to follow these trajectories, will become a stochastic (random) field; see e.g., \cite{JV2019} and references therein. It is possible, however, to generalize the standard flux freezing to stochastic fields in turbulence using a little more advanced mathematics. The result, called stochastic flux freezing developed by \cite{Eyink2011}, states that magnetic field will follow the random particle trajectories in a statistical sense; see also \cite{Eyinketal2013, Eyink2015, Eyink2018, JV2019, JVV2019, Jafari2020}.

\subsection{Stochastic Reconnection}\label{sturbulence2}
Reconnection rate in a laminar flow can be estimated, or defined, in terms of normal diffusion of the magnetic field by magnetic diffusivity $\eta$ on large scales or, in other words, in terms of the (root-mean-square henceforth rms) average distance $\delta(t)$ the field spreads relative to a fixed point. This is of course the Taylor or normal diffusion in which the rms distance between the diffusing material and a fixed point increases as $\delta\sim t^{1/2}$ with time; see e.g., \cite{Eyinketal2013, JVV2019}. For a diffusing magnetic field, $\delta^2\simeq \eta t$. In the absence of turbulence, in a reconnection zone of width $\delta$ and length $\Delta$ (parallel to the local magnetic field), using the Alfv\'en time scale $t_A=\Delta/V_A$, and using mass conservation $ V_A \delta= V_R\Delta$, we recover the reconnection speed;

$$V_R\simeq \Big(\eta {V_A/\Delta}\Big)^{1/2}.$$ 

\begin{figure}[t]
 \begin{centering}
\includegraphics[scale=.38]{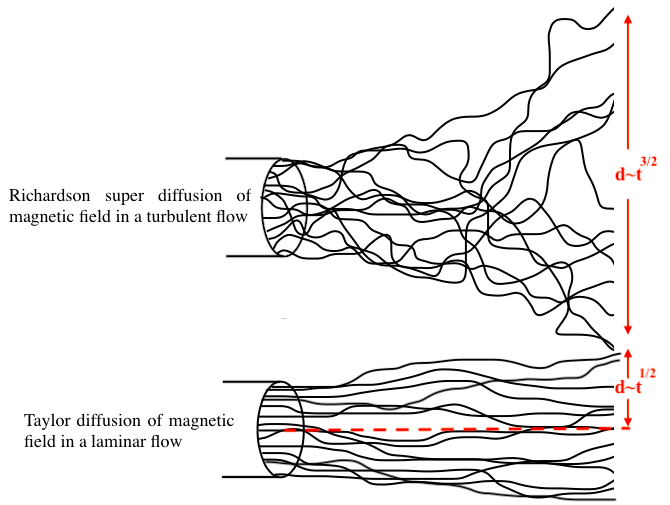}
\caption {\footnotesize {Top: Richardson super diffusion occurs in the turbulence inertial range. The rms width of a bundle of magnetic field lines increases super-linearly with time $\sim t^{3/2}$ ($2$-particle diffusion). Magnetic field follows the flow in a statistical sense hence turbulence creates local current sheets by tangling the magnetic field. The reconnecting small scale fields at multiple local current sheets diffuse super-linearly to larger scales and may give rise to a global reconnection event---stochastic reconnection. Turbulence also increases the spatial complexity of the field which is statistically frozen into the fluid; see \S\ref{SHeating1}. Reconnection relaxes the field decreasing its complexity level. Bottom: Taylor (normal) diffusion occurs in laminar flows, or at scales much larger than the turbulence inertial range. The rms distance of magnetic field lines from a fixed point increases sub-linearly with time $\sim t^{1/2}$ ($1$-particle diffusion). The Sweet-Parker model corresponds to Taylor diffusion. }}\label{zap0}
\end{centering}
\end{figure}

This is, of course, the well-known Sweet-Parker reconnection rate \citep{Parker1957, Sweet1958}.  Reconnection, and/or other instabilities such as tearing modes \citep*{Furthetal.1963}, will in general generate turbulence \citep{Eastwoodetal.2009, JV2018Review}, with the implication that the laminar Sweet-Parker model is far from realistic in turbulent systems such as most astrophysical fluids. In the turbulence inertial range, i.e., at scales larger than dissipative scale but smaller than the larger scales where Taylor (normal) diffusion occurs, diffusing particles will undergo super-linear Richardson diffusion; $d^2\propto t^3$ which is a 2-particle diffusion, i.e., $d$ is the rms separation between any pair of particles undergoing diffusion in the inertial range. If we consider magnetic diffusion in the turbulence inertial range, we have to consider Richardson diffusion of the field, in terms of the rms distance the field spreads during the time $t$; see Fig.(\ref{zap0}). The eddy turnover time, in the inertial range, is of order $t \sim \epsilon^{-1/3}d^{2/3}$ with $d$ being the length scale perpendicular to the mean magnetic field. Here, $\epsilon\simeq V_T^2 V_A/l_\parallel$ denotes the energy transfer rate, with turbulent velocity $V_T$ and parallel energy injection length scale $l_\parallel$. This corresponds to the Richardson diffusion; $d^2\simeq \epsilon t^3$. The super-linear nature of Richardson diffusion broadens the reconnection zone and thereby enhances the reconnection rate. To see this, using mass conservation $ V_Ad= V_R\Delta$, and substituting the Alfv\'en time $t_A=\Delta/V_A$, one arrives at the fast reconnection speed  \citep{LazarianandVishniac1999, Jafarietal2018, Review2019, Review2020}; 

\begin{equation}\label{LV99}
V_R\sim V_T \; \min \Big[ \Big( {\Delta\over l_\parallel}\Big)^{1/2}, \Big( {l_\parallel\over \Delta}\Big)^{1/2}\Big].
\end{equation}

Depending on the parallel (with respect to the local field) length scale of the current sheet, i.e., $\Delta$, and the parallel energy injection scale $l_\parallel$, the smaller ratio, either $(\Delta/l_\parallel)^{1/2}$ or $(l_\parallel/\Delta)^{1/2}$, should be taken in the above formula. This reconnection speed is of order the large turbulent eddy velocity $V_T$; is independent of diffusivity and is in agreement with numerical simulations to date \citep{Kowal2009, Kowal2012}. The stochastic model of reconnection was also examined with a large viscosity to diffusivity ratio in a recent work \citep{Jafarietal2018, JV2018Review}.

\section{Stochastic Heating}\label{SHeating}

In this section, we present the main results of this paper. First, in \S\ref{SHeating1}, employing the recent statistical formalism developed by \cite{JV2019}, we argue that local magnetic reversals are ubiquitous in MHD turbulence, which continuously convert magnetic energy into kinetic and thermal energy; see also \cite{Jafari2020}. Then, we use simple scaling laws of MHD turbulence, in \S\ref{SHeating2}, to support the idea that these local events are efficient in maintaining the turbulent cascade in the inertial range and heat generation at smaller scales down the inertial range. Finally, in \S\ref{SHeating3}, we numerically test our theoretical prediction that magnetic complexity's rate of change should be statistically correlated with magnetic energy dissipation rate $\eta|\nabla{\bf B}|^2$ and/or the rate of change of the kinetic energy.

\subsection{Statistics of Local Reversals}\label{SHeating1}

A fluid parcel of an arbitrary size $l$ located at spacetime point $({\bf x}, t)$ has an average velocity 

\begin{equation}\label{coarsegrain1}
{\bf{u}}_l ({\bf{x}}, t)=\int_V G\Big({{\bf r}\over l}\Big)  {\bf u}({\bf{x+r}}, t) {d^3r\over l^3},
\end{equation}
where $G({\bf r}) = G(r) $ is a smooth and rapidly decaying kernel\footnote{For simplicity, one may also assume $G({\bf{r}})\geq 0$, $\lim_{|\bf r|\rightarrow \infty} G({\bf{r}})\rightarrow 0$, $\int_V d^3r G({\bf{r}})=1$, $\int_V d^3r \; {\bf{r}}\;G({\bf{r}})=0$, $\int_V d^3r |{\bf{r}}|^2 \;G({\bf{r}})= 1$ and $G({\bf{r}})=G(r)$ with $|{\bf{r}}|=r$. }, e.g., $G({\bf r}/l)\sim e^{-r^2/l^2}$ . The real, mathematical field $\bf u$ is sometimes called the bare field while ${\bf u}_l$ is called the renormalized, or coarse-grained, field at scale $l$. This sort of averaging, coarse-graining or renormalizing is in fact the common method using which we usually obtain fluid equations or even the wave equation for a string, i.e., by approximating the average discrete displacements of individual atoms ${\bf x}_i$ by a continuous position function ${\bf x}$, introducing of which remarkably simplifies the calculations. In doing so, we ignore, within a good approximation, that matter is indeed discrete and made of atoms. In a similar way, instead of the vector field ${\bf u (x}, t)$ which mathematically assigns a unique velocity to the point $({\bf x}, t)$ in space and time, we consider the average velocity of a fluid parcel of size $l$ located at $({\bf x}, t)$. Hence, we ignore the fact that the parcel itself is made of many particles which are below our resolution scale $l$. In other words, we look at the fluid with our spectacles off in the sense that we cannot observe or resolve the scales smaller than $l$. Even in quantum field theories, the introduction of such a cut-off scale $l$ is necessary in order to avoid infinite quantities (regularization and renormalization), at least until some day we get a complete theory of nature valid on all scales down to the Planck scale.

\begin{figure}[ht]
 \begin{centering}
\includegraphics[scale=.4]{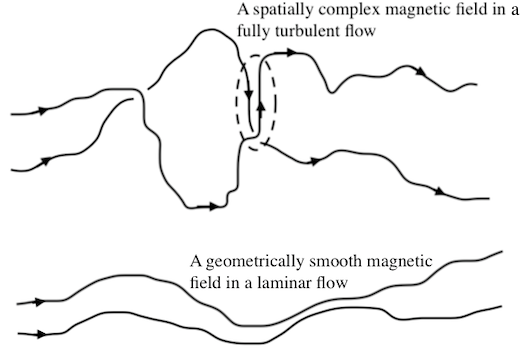}
\caption {\footnotesize {Magnetic complexity, a geometric measure of the spatial complexity of the magnetic field, increases by turbulence as the field is statistically frozen into the turbulent flow. This creates local magnetic shears which give rise to local reconnection events. The final result is the conversion of magnetic energy into kinetic or thermal energy, hence, the rate of change of magnetic complexity is expected to be related to that of magnetic dissipation and kinetic energy.}}\label{complexity}
\end{centering}
\end{figure}

The interaction between a turbulent fluid and the threading magnetic field can be understood in terms of stochastic flux freezing and the spatial complexity of the velocity and magnetic fields. A simplified picture can be described as follows \citep{JV2019, JVV2019, Jafari2020}: 

\textbf{(i) Stochastic flux freezing.} The magnetic field will tend to become increasingly tangled as it statistically follows the turbulent flow. This is an implication of stochastic flux freezing \citep{Eyink2011} on a range of inertial scales $[l,L]$ with $L>l$. The magnetic spatial complexity, quantified by the function \footnote{The motivation behind the definitions (\ref{magnetic}) and (\ref{kinetic}) is as follows: the renormalized field ${\bf u}_l({\bf x}, t)$ represents the average velocity of a fluid parcel of size $l$ at $\bf x$. Because $G({\bf r}/l)$ is a rapidly decaying function, the integral ${\bf u}_l({\bf{x}}, t)=l^{-3}\int_V G({\bf r}/l)  {\bf u}({\bf{x+r}}, t) d^3r$ gets much smaller contributions from distant points at located at $x\gg l$. The large scale field ${\bf u})L$ with $L\gg l$ is the average velocity field of a fluid parcel of scale $L$ at the same point $\bf x$. In a laminar flow whose velocity field has a large curvature radius $\gg L$, we expect $\hat{\bf u}_l.\hat{\bf u}_L\simeq 1$. For a stochastic velocity field in a fully turbulent medium, however, $-1\leq \hat{\bf u}_l.\hat{\bf u}_L\leq 1$ becomes a rapidly varying stochastic variable. This quantity thus measures the spatial complexity (or stochasticity level) of $\bf u$ at point $\bf x$. Its root-mean-square (rms) value tells us how spatially complex (or stochastic) the velocity field is on average in a given volume $V$. In order to obtain a non-negative global quantity, we can volume average ${1\over 2}|\hat{\bf u}_l({\bf x}, t).\hat{\bf u}_L({\bf x}, t)-1|$. Magnetic complexity is defined similarly.}

\begin{eqnarray}\label{magnetic}
S_m(t)={1\over 2}( \hat{\bf B}_l.\hat{\bf B}_L-1)_{rms}
={1\over 2} \Big(  \int_V|\hat{\bf B}_l.\hat{\bf B}_L-1|^2  {d^3x\over V} \Big)^{1/2}
\end{eqnarray}

with unit direction vector $\hat{\bf B}_l\equiv {\bf B}_l/B_l$, will increase over time by the turbulent motions until it reaches a maximum, at which point magnetic tension forces are strong enough to resist further tangling of the field; see Fig.(\ref{complexity}). 

\begin{figure}[ht]
 \begin{centering}
\includegraphics[scale=.22]{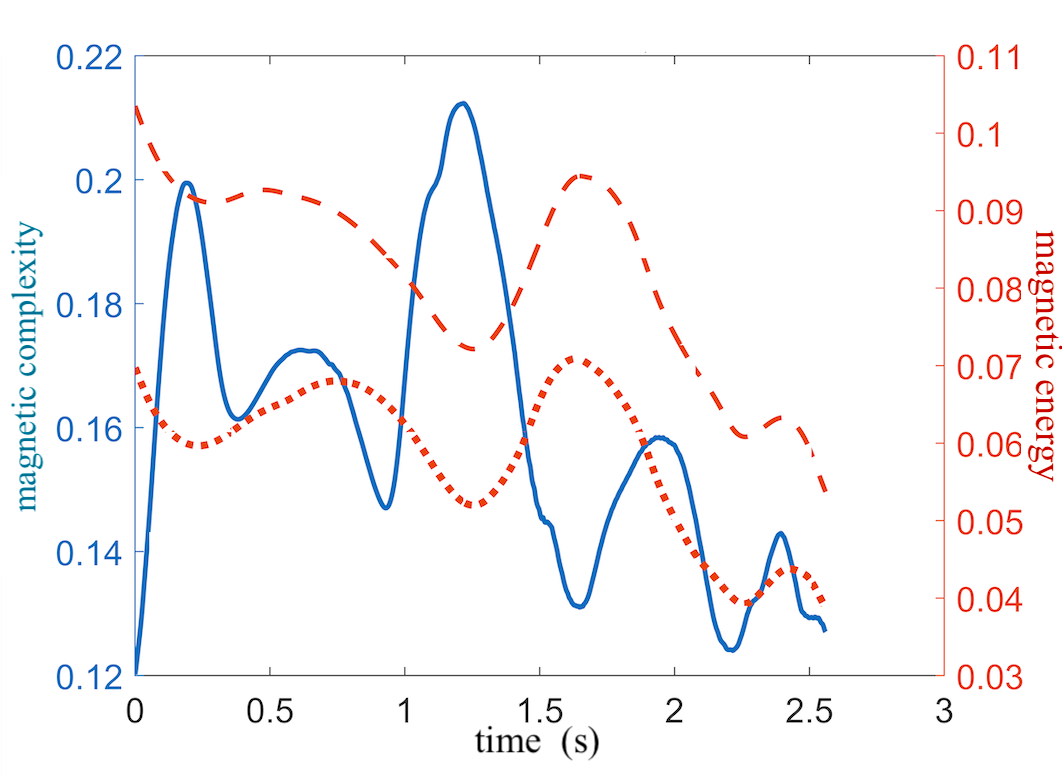}
\caption {\footnotesize {Magnetic complexity $S_m(t)$ (blue, solid curve), cross energy $E_m(t)$ (red, dotted curve) and rms magnetic energy density $(B^2/2)_{rms}$ (red, dashed curve) in a typical sub-volume of the simulation box \citep{Jafari2020}. The magnetic cross energy and rms energy densities follow a similar trend while their trend, i.e., besides small scale fluctuations, shows an anti-correlation with the magnetic complexity $S_m$. One implication is that higher magnetic complexities are associated with magnetic energy conversion.}}\label{zap1}
\end{centering}
\end{figure}

\textbf{(ii) Field-fluid slippage.} Magnetic complexity increases until the tension forces associated with large magnetic field curvatures become strong enough to make the field suddenly slip through the fluid \citep{Eyink2015, JV2019}. Such field-fluid slippages should be a ubiquitous phenomenon in astrophysical systems in which magnetic fields, although typically very complex geometrically, could have survived for millions of years without being infinitely tangled. Filed-fluid slippage corresponds to a sudden drop in magnetic complexity $S_m$ after it reaches a maximum. The corresponding magnetic cross-energy, which is defined as the geometric mean

\begin{equation}\label{Mcrossenergy}
E_m(t)=\sqrt{{B_l^2\over 2} {B_L^2\over 2}    }={1\over 2}B_lB_L,
\end{equation}
will decrease (increase) as the magnetic complexity $S_m(t)$ increases (decreases). In passing, note that the magnetic complexity $S_m$ and cross-energy $E_m$, respectively given by (\ref{magnetic}) and (\ref{Mcrossenergy}), define a scalar field $\psi(t)={1\over 2}{\bf B}_l.{\bf B}_L=({1\over 2}B_l B_L) (\hat{\bf B}_l.\hat{\bf B}_L)$, called scale-split magnetic energy density \citep{JV2019}. Fig.(\ref{zap1}) plots such a typical relationship between $S_m$ and $E_m$ (and also the rms magnetic energy density $(B^2/2)_{rms}$) in a typical sub-volume of the simulation box of an incompressible, homogeneous numerical simulation \citep{Jafari2020}. The anti-correlation between magnetic spatial complexity and magnetic energy density implies that in a fully developed turbulence, the higher the magnetic complexity, the more efficient magnetic energy conversion.

\textbf{(iii) Local reversals.} If the field-fluid slippage is strong enough such that the relaxing field accelerates fluid elements efficiently, converting magnetic energy into kinetic energy, the resultant eruptive, spontaneous fluid motions will increase the spatial complexity of the velocity field, which is defined by

\begin{eqnarray}\label{kinetic}
S_k(t)={1\over 2} (\hat{\bf u}_l.\hat{\bf u}_L-1)_{rms}.
\end{eqnarray}

Thus $\partial_t S_k(t)$ will take positive values as $S_m(t)$ reaches its maxima (at times for which $\partial_t S_m=0$ \& $\partial_t^2 S_m<0$).

\begin{figure}[h]
 \begin{centering}
\includegraphics[scale=.57]{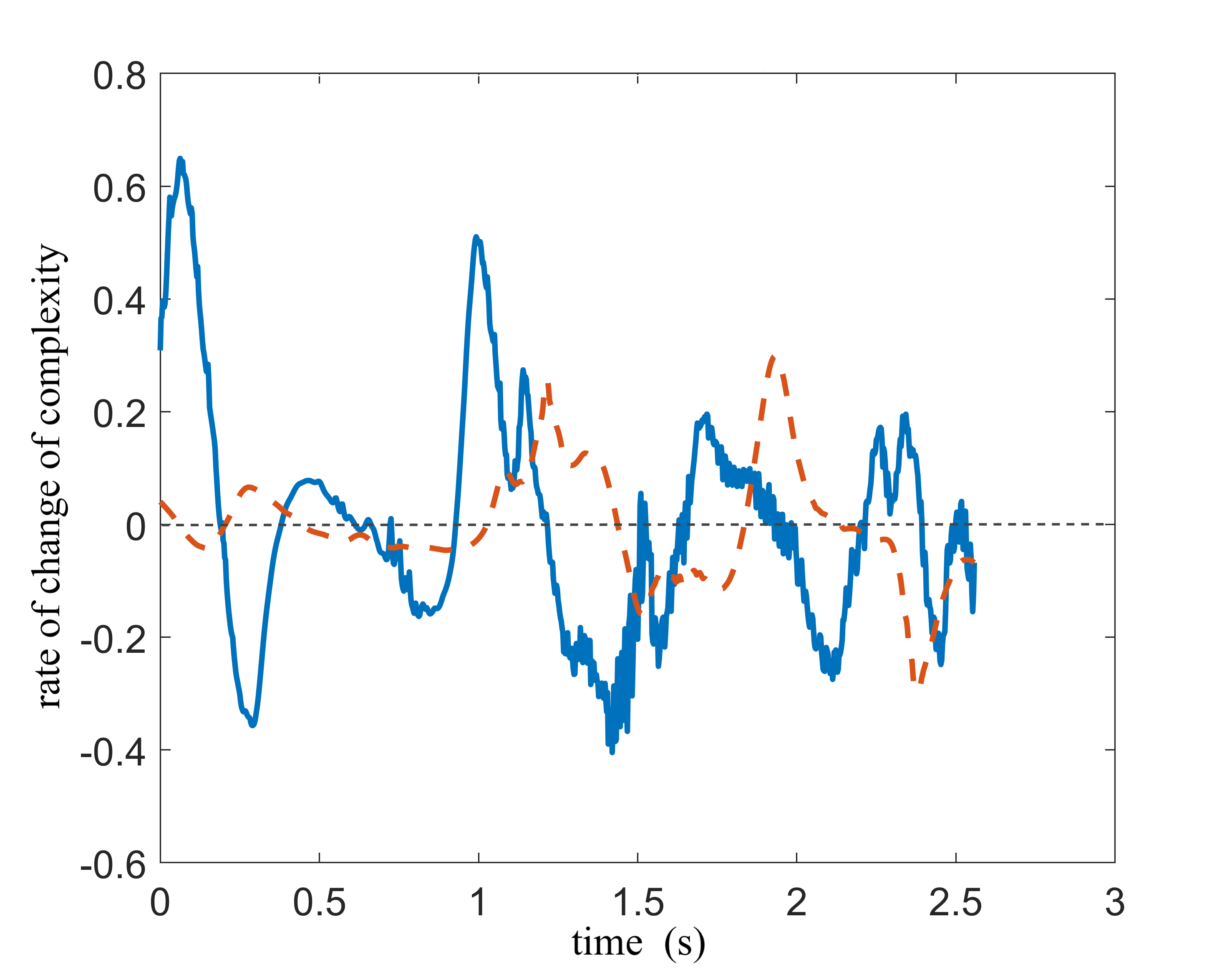}
\caption {\footnotesize {The rate of change of the magnetic (blue, solid curve) and kinetic (red, dashed curve) , complexities, $\partial_t S_m$ and $\partial_t S_k$, in the same sub-volume used in Fig.(\ref{zap1}). At points where $\partial_t S_m=0\;\&\; \partial_t^2 S_m<0$, i.e., where the solid blue curve vanishes with a negative slope, the magnetic complexity reaches a local maximum and magnetic reconnection peaks. As magnetic complexity starts to decrease, we have $\partial_t S_m<0$ and the reconnecting field pushes the fluid and increases the kinetic complexity; $\partial_t S_k>0$ \citep{Jafari2020}. During a field-fluid slippage, $S_m(t)$ reaches a maximum while $S_k(t)$ is not affected.}}
\label{dphiT1}
\end{centering}
\end{figure}

\begin{figure}[h]
 \begin{centering}
\includegraphics[scale=.59]{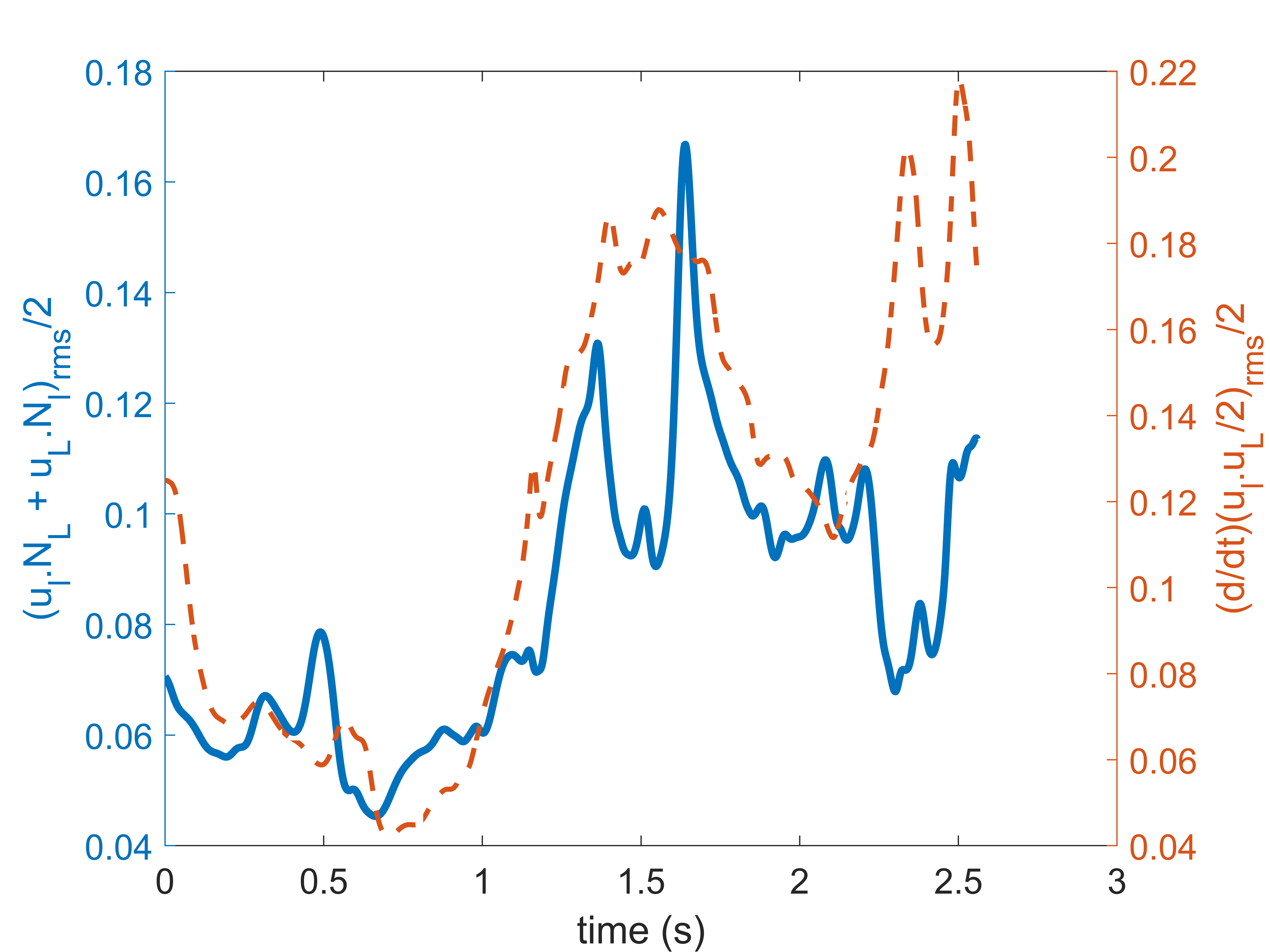}
\caption {\footnotesize {Reconnection power ${\cal P}$ (solid, blue curve) and $(\partial_t\Psi_{lL})_{rms}$ (dashed, red curve) in the same region of the simulation box as the one in Fig.(\ref{dphiT1}). These two functions are strongly correlated (with a typical cross-correlation above $0.90$) in regions where magnetic reversals seem to be strong \citep{Jafari2020b}. }}\label{zap10}
\end{centering}
\end{figure}

Fig.(\ref{dphiT1}) illustrates this typical behavior between $\partial_t S_m$ and $\partial_t S_k$ in the same sub-volume as in Fig.(\ref{zap1}). The corresponding cross-energy is defined as 

$$E_k(t)=\sqrt{{u_l^2\over 2} {u_L^2\over 2}    }={1\over 2}u_lu_L,$$

which along with the kinetic complexity (\ref{kinetic}) define the scale-split kinetic energy density  $\Psi(t)={1\over 2}{\bf u}_l.{\bf u}_L=({1\over 2}u_l u_L) (\hat{\bf u}_l.\hat{\bf u}_L)$ \citep{JV2019}. Incidentally, note that the acceleration of fluid particles by a slipping magnetic field ultimately results from Lorentz forces ${\bf N}_l=({\bf j\times B})_l-{\bf j}_l\times {\bf B}_l$ with electric current $\bf j$, acting on the fluid elements at an arbitrary scale $l$. The reconnection power on an arbitrary range of inertial scales $[l, L]$,  defined as ${\cal P}={1\over 2} ({\bf u}_l.{\bf N}_L+{\bf u}_L.{\bf N}_l)_{rms}$, therefore, is expected to be strongly correlated with the rate of change of the kinetic energy $(\partial_t\Psi)_{rms} $ \citep{Jafari2020b}; see Fig.(\ref{zap10}).

In short, on the one hand, the interplay between turbulence and magnetic field results in rapid temporal variations in magnetic complexity $S_m$ in an arbitrary spatial volume $V$. A sudden decrease in $S_m$ indicates the presence of field-fluid slippage and/or local magnetic reversals, which are indeed observed almost everywhere in the inertial range of MHD turbulence \citep{Eyinketal2013, Eyink2015, Jafari2020}. On the other hand, the magnetic complexity is anti-correlated with magnetic energy \citep{JV2019, Jafari2020}, therefore, these ubiquitous local reversals imply magnetic energy conversion. The range of scales $[l, L]$ is arbitrary in the above arguments. Hence, at larger scales in the inertial range, these reversals will in general enhance turbulent diffusion whereas at the smaller scales, they will enhance the heating process in the dissipative range. All in all, this picture suggests that the magnetic field in a turbulent fluid will be spatially complex, in the sense that there will exist intense local magnetic shears which either annihilate the magnetic energy and/or cause local reversals. In fact, the diffusion of these small scale effects by means of superlinear Richardson diffusion in turbulence \citep{JVV2019} can lead to a global reconnection event---stochastic reconnection \citep{LazarianandVishniac1999, Eyinketal2013}. In the next section, we use scaling laws in MHD turbulence to show that these reversals are indeed efficient in both maintaining the turbulence in an arbitrary inertial scale $l$ and also enhancing the magnetic-to-thermal energy conversion at smaller scales. 

\subsection{Scaling Laws and Local Reversals}\label{SHeating2}

In the preceding subsection, based on analytical and numerical arguments, we reasoned that the local reversals occur frequently in MHD turbulence. In this subsection, we use conventional scaling laws in MHD turbulence to support our previous arguments, and also to show the efficiency of local reversals in enhancing the turbulence and heating the medium. For simplicity, we will use the Goldreich-Sridhar model \citep{GoldreichandSridhar1995, GoldreichandSridhar1997} of MHD turbulence, however, our arguments are model-independent and quite general.

Let us consider a local reconnection event to see how it interacts with the local turbulent cascade. Suppose that turbulence is generated by energy injection at some scale $l$, which creates an rms turbulent velocity $V_T$ at the largest scales of the ensuing cascade. This energy can be injected by e.g., a global reconnection, or the source may be external. In any case, the kinetic energy of large scale motions, $V_T^2$, will be much larger than that associated with any smaller scale $k^{-1}$, which we denote by $v_k^2$. Thus, in order to enhance, or sustain for that matter, the turbulence at a scale $\lambda<l$, a local reconnection event would only have to inject a small amount of energy $\lesssim v_\lambda^2$, much smaller than that contained at larger scales, $v_\lambda^2 < V_T^2$. The available energy for this local event comes from the local magnetic energy, $b^2$. Since the local mean magnetic energy, unlike the kinetic energy, depends only slightly on scale, a local reconnection has enough magnetic energy to provide the local turbulence with a kinetic energy comparable to the turbulent energy at that scale. Stochastic reconnection feeds turbulence at all scales.

In order to quantify the above argument, we start by noting that MHD turbulence is anisotropic in general. Suppose energy is injected at a scale $l_\parallel$, parallel to the mean magnetic field, with corresponding perpendicular scale $l_\perp=l_\parallel (V_T/V_A)$, which creates the r.m.s turbulent velocity $V_T$ at this scale \footnote{An implicit assumption here is that the magnetic diffusivity is of the same order as the viscosity, i.e., a magnetic Prandtl number of order unity. This is to assure that the viscous damping scale, below which hydrodynamic motions but not necessarily magnetic structures are dissipated, is of order the resistive dissipation scale, below which magnetic field is dissipated.}. The non-linear timescale in Goldreich-Sridhar \citep{GoldreichandSridhar1995} MHD turbulence is given by $\tau_{nl} \simeq k_{\parallel}V_A/(k_{\bot}^2 v_k^2)$ with the r.m.s eddy velocity $v_k$ (which is clearly scale-dependent as its subscript indicates). The critical balance condition on all scales, $k_\| V_A\approx k_\perp v_k$, then leads to the energy transfer rate 

\begin{equation}\label{GenRate}
\epsilon\simeq {V_T^2 \over (l_\parallel/V_A)}\simeq {v_k^2\over \tau_{nl}  }\simeq k_\perp v_k^3.
\end{equation}

 The assumption of constant energy transfer rate \citep{Kolmogorov1941}, $\epsilon\simeq k_\perp v_k^3\simeq const.$, leads to $v_k\propto k_\perp^{-1/3}$ corresponding to the famous, Kolmogorov-type, energy power spectrum $E_{GS}(k_\perp)\sim k_\perp^{-5/3}$ in Goldreich-Sridhar MHD turbulence model. Putting this together, we find 

\begin{equation}\label{v_k}
v_k \simeq  V_T\left({k_\perp l_\parallel V_T\over V_A}\right)^{-1/3},
\end{equation}

and 

\begin{equation}\label{67}
k_{\parallel} \simeq l_\parallel ^{-1}\left({k_\perp l_\parallel V_T\over V_A}\right)^{2/3}.
\end{equation}
 
Let us focus on a local reconnection at a parallel scale $k_{\parallel}^{-1}<l_\parallel$, which is still much larger than the dissipation scale. Mass conservation leads to a local reconnection speed of order $v_R\simeq V_A(k_\parallel/k_\perp)$. This is of order the local turbulent velocity $v_k$, if we use the critical balance condition in Goldreich-Sridhar model \citep{GoldreichandSridhar1995}; $v_R\sim v_k$. The ejection velocity $v_e$ will be in general larger than the local reconnection velocity, $v_e>v_R$ since $v_e/v_R\simeq k_\perp/k_\parallel >1$. This is how a local stochastic reconnection enhances the local turbulent cascade.

At smaller scales, where resistivity derives reconnection, the local reconnection speed scales as $v_R\sim \eta k_\perp$. Using $v_R\simeq V_A(k_\parallel/k_\perp)$, the local reconnection speed is of order

\begin{equation}
v_R\sim V_T^{1/2}\Big( {\eta V_A\over l_\parallel}\Big)^{1/4}.
\end{equation}
Note that this is basically the local r.m.s turbulent eddy velocity $v_k$ given by eq.(\ref{v_k}) with $k_\perp^{-1}\simeq \eta/v_R$ as the outflow width, which is set by the resistivity. The largest perpendicular wavenumber in the turbulent cascade is given by 

\begin{equation}
k_\perp\sim {v_R\over \eta}\propto \eta^{-3/4}.
\end{equation}

At wavenumbers larger than this, reconnection will generate its own local turbulence \citep{LazarianandVishniac1999}. This means at these small scales, local reconnections interact with the turbulence; they may generate a local cascade or enhance the exiting turbulence. This in turn enhances particle diffusion in the medium. Turbulent diffusion (Richardson diffusion) is super-linear and thus much more efficient than normal (linear) diffusion \citep{JVV2019}. In the absence of turbulence, the rms particle separation would be governed by much slower normal diffusion.

Let us compare the global magnetic heating with the turbulent heating in stochastic reconnection. Consider a fully turbulent reconnection zone, of spatial size $\Delta$ and with an r.m.s turbulent velocity $V_T$ at large scales, embedded in large scale field $B$. The energy dissipation rate at the reconnection zone is $\bf{J.E}$, with current $J=|\nabla\times{\bf{B}}|\simeq B/\Delta$ and electric field $E\simeq V_R B$ where $V_R$ is the reconnection speed. Consequently, the magnetic energy dissipation rate is roughly of order

\begin{equation}
\epsilon_b\simeq B^2 {V_R\over \Delta}\simeq B^2 {V_T\over (l_\parallel \Delta)^{1/2}},
\end{equation}
where we have used eq.(\ref{LV99}) for the reconnection speed. This is basically $\epsilon_b\simeq B^2 /\tau_R$ with reconnection rate $\tau_R\simeq \Delta/V_R$. On the other hand, the kinetic energy dissipation rate in sub-Alfv\'enic turbulence \citep{GoldreichandSridhar1995, LazarianandVishniac1999, JV2018Review} scales as 
  
 \begin{equation}
 \epsilon_v\simeq {V_T^2\over (l_\parallel/V_A)}.
 \end{equation}
Because we expect $V_T/V_A<(l_\parallel/\Delta)^{1/2}$, in the presence of a large scale magnetic field reversal, i.e., a global reconnection event, the magnetic heating rate is larger or at least of order the turbulent heating rate $\epsilon_b \gtrsim \epsilon_v$. Thus, a global reconnection in a turbulent medium might enhance the continuous heat generation by the turbulence and Alfv\'en wave dissipation by the enhanced turbulence can increase this to even higher rates.

What is the heating rate associated with a local stochastic reconnection event above the dissipation scale? The magnetic heating rate is estimated as
\begin{equation}
\epsilon_{b, loc}\sim b^2 (k_\parallel v_R) \sim b^2  ( v_kk_\perp ).
\end{equation}

And, the turbulent energy dissipation rate, from eq.(\ref{GenRate}), is
\begin{equation}
\epsilon_{v, loc}=\epsilon_v \sim v_k^2 (v_k  k_\perp ) .
\end{equation}
Magnetic heating rate associated with a local event, above the dissipation scale, is at least as efficient as the turbulent heating.  A  larger ratio of viscosity to resistivity than unity which is assumed here, leads to more complications. We do not consider this regime here.

 In conventional reconnection models, the kinetic energy injected into the medium at large scales, during a global reconnection event, has a long way to reach down the dissipation scale, which simply means long time scales. Above, we argued that this energy injection may enhance diffusion at large scales without directly affecting small scale turbulent motions. Closely related is the notion that, because of the long time scales involved, the final conversion of this kinetic energy into thermal energy may happen long after a global reconnection ceases. In contrast, with energy injection at all scales in stochastic model, a considerable fraction of kinetic energy should be converted into heat at shorter time scales, i.e., during a global reconnection. The democratic participation of all scales in stochastic reconnection, therefore, translates into a faster furnace down the cascade. Super-linear (Richardson) diffusion broadens the outflow zone, during a reconnection event, and increases the flux of the ejected matter. Since ejection velocity and reconnection zone's length are almost fixed observables in all models, and since stochastic reconnection is fast, this implies more efficient kinetic energy injection into the turbulent cascade.

\subsection{Energy Conversion and Local Reversals }\label{SHeating3}

\begin{figure}[h]
 \begin{centering}
\includegraphics[scale=.22]{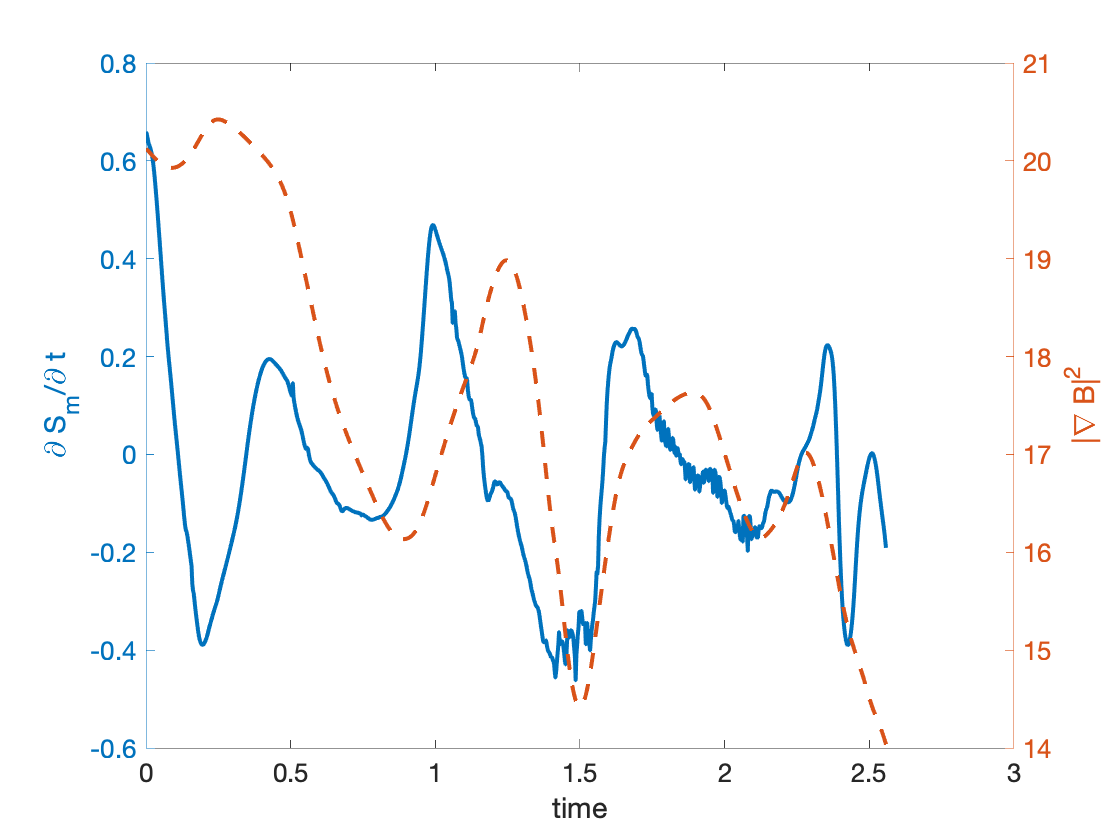}
\caption {\footnotesize {Rate of change of magnetic complexity $\partial_t S_m$ (solid, blue curve) and the rate of change of the (rms) magnetic energy dissipation $\propto |\nabla {\bf B}|^2$ in a randomly selected region of the simulation box. As magnetic complexity increases (decreases), magnetic shears and consequently magnetic dissipation rate are expected to be enhanced (declined) in a statistical sense. }}\label{zap12}
\end{centering}
\end{figure}

\begin{figure}[h]
\begin{centering}
\includegraphics[scale=.22]{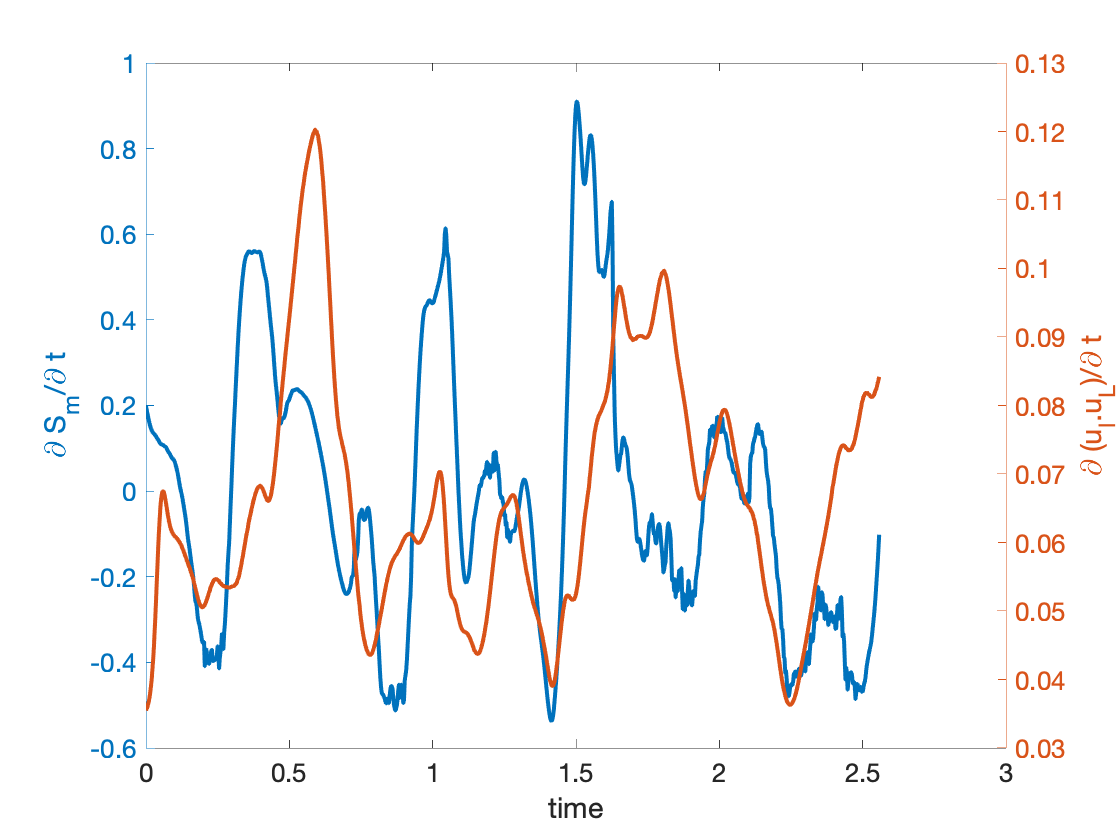}
\caption{\footnotesize{Rate of change of magnetic complexity $\partial_t S_m$ (solid, blue curve) and the rate of change of the (rms) kinetic energy $\propto \partial_t({\bf u}_l.{\bf u}_L/2)_{rms}$, whose behavior mimics that of $\partial_t (u^2/2)$ \citep{Jafari2020}, in a randomly selected sub-volume where magnetic dissipation seems to be less correlated with $\partial_t S_m$. Instead, $\partial_t S_m$ shows a strong correlation with the (rms) rate of change of kinetic energy $\partial_t(u^2/2)$. Here, unlike Fig.(\ref{zap12}), the magnetic energy seems to be mostly converted to kinetic energy rather than thermal energy at smaller scales.}}\label{zap13}
\end{centering}
\end{figure}

In the preceding subsections, we used analytical and numerical results from previous work as well as conventional scaling laws of MHD turbulence to illustrate the efficiency of local reversals in enhancing turbulence and heat generation. In this subsection, we test these results numerically by looking at the correlation between magnetic energy dissipation rate $\epsilon_m(t)=\eta |\nabla {\bf B}|^2$ and the rate of change of the magnetic complexity, $\partial_tS_m(t)$. We use the homogeneous, incompressible MHD numerical simulation archived in the online, web-accessible database of Johns Hopkins University\footnote{Forced MHD Turbulence Dataset, Johns Hopkins Turbulence Databases, https://doi.org/10.7281/T1930RBS (2008).} \citep{JHTB1, JHTB2}. This is a direct numerical simulation (DNS), using $1024^3$ nodes, which solves incompressible MHD equations using pseudo-spectral method. The simulation time is $2.56$ and $1024$ time-steps are available (the frames are stored at every $10$ time-steps of the DNS). Energy is injected using a Taylor-Green flow stirring force. We divide up the simulation box into sub-volumes with randomly selected coordinates and sizes in order to obtain a larger number of statistical samples. Figures (\ref{zap12}) and (\ref{zap13}), for example, are produced in such randomly selected regions of the box. The typical size of these boxes is $\gtrsim 10^{4-5}$ and the scales $l$ and $L$ take typical values $3\leq l< L\leq 21$ (in grid units).

Fig.(\ref{zap12}) plots both the rate of change of magnetic complexity $\partial_tS_m$ and magnetic energy dissipation rate $\propto |\nabla {\bf B}|^2$ in one randomly selected sub-volume of the simulation box far away from the region considered in Fig.(\ref{zap12}). A strong correlation is observed between these quantities on average (a cross-correlation about $0.6$). In this case, it seems that we are dealing with a region in the simulation box where instead of magnetic reversals, the small scale local magnetic field gradients annihilate the field converting magnetic energy mostly to thermal energy, as in the nanoflare theory of Parker. In regions where magnetic energy dissipation is not strongly correlated with the rate at which the magnetic complexity changes, the latter is usually correlated with the rate at which kinetic energy changes, i.e., $\partial_t ({\bf u}_l.{\bf u}_L/2)$. This quantity is important in the considerations related to magnetic reconnection, but in any case, its trend over time closely resembles that of $\partial_t(u^2/2)$ (the same argument applies to magnetic field too, see Fig.(\ref{zap1}) in \S\ref{SHeating1}). Therefore, our qualitative discussion here is not sensitive to this choice; see also \cite{Jafari2020}. Fig.(\ref{zap13}) plots the rate of change of magnetic complexity $\partial_tS_m$ and $\partial_t ({\bf u}_l.{\bf u}_L/2)$. In this region, unlike the region corresponding to Fig.(\ref{zap12}), the change in magnetic complexity shows strong correlation with the change in kinetic energy, suggesting magnetic to kinetic, rather than magnetic to thermal, energy conversion.

We should emphasize that the correlations between different quantities discussed in this section, such as the rate of change in magnetic complexity and magnetic dissipation rate, should be understood as statistical cross-correlations between time series constructed from randomly selected samples in a simulation box, therefore, they are meaningful only in a statistical sense in terms of the trends of these time series. In fact, more detailed numerical analyses based on a larger number of samples, i.e., sub-volumes of the simulation box or even independent runs, are required to carefully test the analytical arguments advanced here. Our short treatment in this section should be regarded only as a self-consistency check rather than such a detailed numerical study.

\section{Summary and Conclusions}\label{Discussion}

In this paper, we have invoked analytical and numerical results from previous work \citep{JV2019, JVV2019, Jafari2020} to argue that ubiquitous local magnetic reversals in MHD turbulence, which by the way play a major role in the stochastic model of magnetic reconnection \citep{LazarianandVishniac1999}, efficiently enhance the turbulence, in the inertial range, and the heat generation, in the dissipative range. Reconnection events seem to be ubiquitous in turbulent environments including the solar corona, therefore, local reversals associated with nanoflares in Parker's theory \citep{Parker1972} and their collective outcome as stochastic reconnection events may at least partly explain the coronal heating problem. The main difficulty with Parker's model lies basically in the detection of individual nanoflares observationally. Our approach here does not of course address this problem directly, however, relating local reversals to the recently formulated notion of magnetic complexity \citep{JV2019} and stochastic reconnection \citep{LazarianandVishniac1999} at larger scales may in fact provide an indirect way to understand the coronal heating process.

Previous work \citep{JV2019, JVV2019, Jafari2020, Jafari2020b} has established a statistical formalism to study the spatial complexity/stochasticity level of a given vector field such as turbulent magnetic and velocity fields, which can be used to study reconnection and small scale magnetic reversals in MHD turbulence. In this picture, magnetic reversals are studied in terms of the time evolution of magnetic and kinetic complexities and energies at arbitrary inertial scales. A Lorentz force responsible for reconenction at any given scale which arises from sub-scale electric currents has also been introduced and its correlation with magnetic complexity has been analytically and numerically studied \citep{Jafari2020b}. Based on these recent developments, in this paper, we have argued that small scale magnetic reversals in the inertial range of turbulence result from tangling of the magnetic field by the turbulent motions. This can be understood in terms of stochastic flux freezing \citep{Eyink2011}, which is a generalization of the conventional flux freezing theorem \citep{Alfven1942} in highly conducting fluids. The more spatially complex the magnetic field becomes by statistically following the turbulent flow, the larger number of small scale current sheets will be present. In these regions, magnetic energy will be converted into heat by direct dissipation, or they will undergo small scale magnetic reversals thereby injecting energy to the flow. As a consequence, we expect that the rate at which magnetic complexity changes, i.e., $\partial_t S_m(t)$, be positively correlated with magnetic dissipation rate $\eta|\nabla{\bf{B}}|^2$ in the former case and with the rate at which the kinetic energy changes $\partial_t u^2/2$ in the latter case. Numerical simulations of incompressible, homogeneous MHD turbulence seem to be in agreement with this picture, although more detailed numerical studies are in demand to establish a firm evidence. We have also backed up our analytical arguments by conventional scaling laws of MHD turbulence to show that small scale reversals are indeed efficient in enhancing the turbulence and heat generation.

All in all, the arguments advanced in this paper suggest that small scale magnetic reversals are ubiquitous in MHD turbulence and may play an important role in heating turbulent and highly magnetized fluids such the solar corona. The other implication is that stochastic magnetic reconnection, which consists of many simultaneous local reversals, is more efficient in enhancing the turbulence and heating the fluid than conventional reconnection models. The statistical picture presented in this paper, based on coarse-grained fields and their spatial complexities, can be regarded as a modern reformulation of the nanoflare theory put forward by \cite{Parker1972}.

\bibliography{NanoflareApJ}{}
\bibliographystyle{aasjournal}



\end{document}